# Transition from Established Stationary Vision of Black Holes to Never-Stationary Gravitational Collapse

Miguel Piñol Ribas, Ignacio López Aylagas

*Abstract*— The established concept of black hole emerged from several results founded on Einstein's General Theory of Relativity. In this article, the relationship between these results is analyzed, and it is pointed out how, in spite of being individually correct, the sum of all them do not actually determine the existence of black holes. Some logical incompatibilities in the standard Black Hole model are put into evidence, and the alternative scheme of the Never-Stationary Gravitational Collapse is defended. To illustrate the essence of the new paradigm, a simple but conceptually complete toy model is worked out and a qualitatively suitable metric for Never-Stationary Gravitational Collapse is presented.

contact e-mail: mpinolri7@alumnes.ub.edu

## I. Introduction

THE concept of black hole has been historically raised from several results in General Relativity. Nevertheless, the identification of certain implications of these results has not been established in the right way, as this article will prove in detail. In first place, we will review the list of these historical results and to point out their actual relevance:

1- In 1916, Karl Schwarzschild found an exact solution for Einstein's gravitational equations, which pretended to describe the field created by a point particle. The solution is also valid for every spherically symmetric body at a distance larger than its radius. [1]

2- In 1939, Oppenheimer and Volkov discovered that there existed an upper limit for the mass of neutron stars, and over this limit gravitational collapse could not be avoided. [2]

3- In 1967, John Wheeler coined the term "black hole" to name a "gravitationally completely collapsed star". [3]

4- In 1970, Stephen Hawking and Roger Penrose demonstrated that under certain circumstances singularities could not be avoided, according to Albert Einstein's General Theory of Relativity. This is known as the Hawking-Penrose theorem of singularity. [4]

These previous results, as we have pointed out, basically proceeded from Einstein's General Relativity. On the other hand, there exist two important issues in black hole description which are not only related to General Relativity but also with Thermodynamics and Quantum Field Theory:

5- In 1972, Jakob Bekenstein defined the entropy of black holes and thermodynamically deduced the need of black hole radiation. [5]

6- In 1974, Stehphen Hawking justified Bekenstein's speculations about black hole radiation from the perspective of Quantum Field Theory. His model implies the creation of particles of negative mass near the horizon of events of black holes. The conservation of information was not clearly assured by this model. [6, 7]

## II. Towards a new paradigm in the understanding of gravitational collapse

The previous points, contrarily to what is widely believed, do not lead to the proof of the existence of black holes. The reasoning in which the existence of black holes is based is the following one:

1- There exist stars which are massive enough to exceed at the end of their "vital cycle" the Oppenheirmer-Volkov limit. Those stars must finally enter collapse.

2- According to the theorem of singularity, all the mass inside an event horizon must reach a single central point

3- The solution for the single point mass particle is the Schwarzschild metric, which describes a black hole.

In this reasoning, there is a weak point: entering collapse does not immediately lead to the formation of an event horizon, and therefore entering collapse does not necessarily lead to a complete collapse, as the theorem of singularity is consequently not properly applicable. Certainly, the period of time involved in the process of collapse may be demonstrated to be infinite from the point of view of an external observer (that is, of any viable scientist).

Supporters of black holes do actually know the fact of the infinite lasting collapse from the perspective of an external observer, but they radically state that the "correct" point of view is that one of the "free falling observer", who in its proper system of reference would measure a finite period of time for the collapse. Then, the external vision is usually justified in the following way: "the free falling body has already got the central singularity, but as the light emitted from the body inside the black hole never gets outside the black hole, we cannot see it to fall; in addition, the light which has been emitted near the horizon of events of the black hole arrives to us with a great delay, giving to us the illusion that it is already falling".

There are reasons of weight in order to reject the previous explanation. One of them, consists in the simple fact that if the body reached the interior of the black hole but the light could not be emitted from there that would imply that the body would have traveled at a speed greater than the speed of light (*See Appendix A*). On the other hand, there is even a more important reason. In Relativity, time and space are elastic, but causal order is always preserved: the order of causally related events is invariant for any system of reference.

Let's see in which manner would an external observer see a body freely falling inside a black hole. The external observer would see the body asymptotically approaching the horizon of events of the black hole, but never to cross it. This fact is not, at the moment, incompatible with the fact that it would cross that surface after all the History of Universe (which would last only for a quite limited period of time from the perspective of the free-falling observer), but in those circumstances the limits of evidence-based Physics would be clearly trespassed.

Nevertheless, much before the end of all the History of Universe, the free-falling observer would have been completely evaporated (before having crossed the event horizon), as ruled by the laws of Thermodynamics. According to Hawking's law for black hole radiation, every black hole must evaporate in a finite period of time. Therefore, the external observer of the falling body will see the falling body to evaporate before having crossed the horizon of events of the black hole: we must conclude that is the causal order, and that should be for any observer.

If we take this as a premise, we will notice that the diverging perspectives of the "free falling observer" and the "external observer" are no longer contradictory: The free falling observer would get the singularity in a finite period of time, but in a lesser period of time it gets evaporated, as it has been fully emitted as radiation.

What we have said about a body falling inside a black hole is also valid for a collapsing body trying to form a black hole. Therefore, collapsing bodies never become black holes, but they asymptotically tend to form a black hole, till the thermal radiation (which must be emitted by entropic reasons, in spite if the mechanism differs from that proposed by Hawking, as we are not in front of an actual black hole) evaporates them before having arrived to their goal. An equivalent thesis has already been defended by Vachaspati, Stojkovic and Krauss. [8, 9]

One objection to this point of view could be that the theorem of singularity would not be accomplished, but as we have pointed out previously the theorem of singularity is not rightly applicable in a collapsing star before the formation of its event horizon, which never takes place. On the other hand, it should be reviewed if the theorem of singularity is valid or not when we work not only with general relativity but also with quantum-thermodynamical effects. Outside the conditions in which it was proven, there is no need to consider it longer valid.

The metric of a collapsing body, thus, will never be that of Schwarzschild, as it is never completely collapsed, but a non-stationary metric. In the following section we develop a toy model in order to obtain an approximation for the metric of collapsing bodies.

### III. TOY MODEL

Let's consider a spherical distribution of matter starting to collapse towards a central point.

Given the spherical symmetry of the problem, we can imagine that the metric of the system will be "Schwarzschild-like", but with an inner mass "$\mu$" depending on the radial distance to the center, since the whole mass is not concentrated in a single point but distributed along the radius of the collapsing star:

$$ds^2 = -\left(1 - \frac{2\mu_{(r,t)}}{r}\right)dt^2 + \left(1 - \frac{2\mu_{(r,t)}}{r}\right)^{-1} dr^2 + r^2 d\Omega^2$$

(eq. 1),

where $\mu_{(r,t)}$ is the total energy-mass limited by a spherical surface of radius "r" at time "t".

In order to determine the characteristics of $\mu_{(r,t)}$, we virtually divide the collapsing body into a series of surfaces, so that every "σ-surface" will delimit a constant mass $\mu_\sigma$. If the body is collapsing, the radius of every surface will change along time (*See Figure 1*).

In order to solve the dynamics of the system easily, but without renouncing to the essential physics of the problem, we will take a couple of approximations. The first one will consist in considering that every element of mass is moving straightforwardly towards the center, with no angular movement; therefore, $d\Omega^2$ will be strictly zero for every element of mass $(d\Omega^2=0)$. We are also going to assume that the collapsing speed of every surface is very close to the speed of the light, so that we can take the approximation $ds^2=0$; this corresponds to the fastest possible collapse, and consequently in any physical situation the collapse will be slower (*See Appendix C*).

According to the previous assumptions, the equations of motion of each spherical surface may be simplified to the following one:

$$0=-\left(1-\frac{2\mu_\sigma}{r_\sigma(t)}\right)dt^2+\left(1-\frac{2\mu_\sigma}{r_\sigma(t)}\right)^{-1}dr_\sigma^2 \quad \text{(eq. 2)},$$

where $\mu_\sigma$ is the mass contained in the space delimited by the "σ-surface". By definition, as the surfaces which we consider are co-moving with the mass, then $\mu_\sigma$ is a constant for every given "σ-surface". Nonetheless, the radius of the "σ-surface" will vary along time according to the previous expression, from which we can deduce the expression for the speed of collapse:

$$\left|\frac{dr_\sigma}{dt}\right|=1-\frac{2\mu_\sigma}{r_\sigma} \quad \text{(eq. 3)}$$

As we are considering a situation of collapse and not of expansion, the sign of the radial movement must be negative:

$$\frac{dr_\sigma}{dt}=-\left[1-\frac{2\mu_\sigma}{r_\sigma}\right] \quad \text{(eq. 4)}$$

By grouping variables:

$$\frac{r_\sigma}{r_\sigma-2\mu_\sigma}dr_\sigma=-dt \quad \text{(eq. 5)}$$

In order to solve the equation "eq. 5", we effectuate the following change of variables:

$$r_\sigma^\Delta=r_\sigma-2\mu_\sigma \quad \text{(eq. 6)}$$

where $r_\sigma^\Delta$ represents the distance of the "σ-surface" to the radius of Schwarzschild corresponding to the mass contained in the space delimited by that surface, that is, $\mu_\sigma$.

With this change of variables, equation "eq. 5" is transformed into:

$$\left(1+\frac{2\mu_\sigma}{r_\sigma^\Delta}\right)dr_\sigma^\Delta=-dt \quad \text{(eq. 7)}$$

By integrating, we obtain:

$$r_\sigma^\Delta+2\mu_\sigma\ln(r_\sigma^\Delta)=-t+A(2\mu_\sigma) \quad \text{(eq. 8)}$$

where $A(2\mu_\sigma)$ is a constant of integration (representing the "contour conditions" of the problem) which may depend only on the concrete "σ-surface" which we are considering; that is, it will be a function of $\mu_\sigma$.

Now let's make some simplifications for large values of t. When $t\gg|A(2\mu_\sigma)|$ the sign of the right side of the equation is negative, and therefore the global sign of the left side of the equation must be as well. In the left side, the minus sign must proceed from the only term in that side which is able to the negative; then, if $\ln(r_\sigma^\Delta)$ is at the same time very great in absolute value but negative, this implies a very small value for $r_\sigma^\Delta$, so that the following approximation is justified:

$$2\mu_\sigma\ln(r_\sigma^\Delta)\approx-t+A(2\mu_\sigma) \quad \text{(eq. 9)}$$

from where we can obtain an expression for the time dependence of $r_\sigma^\Delta$:

$$r_\sigma^\Delta\approx e^{\frac{-t+A(2\mu_\sigma)}{2\mu_\sigma}}=e^{\frac{A(2\mu_\sigma)}{2\mu_\sigma}}e^{\frac{-t}{2\mu_\sigma}} \quad \text{(eq. 10)}$$

Thus, the distance $r_\sigma^\Delta$ asymptotically approaches zero, and the radius of the "σ-surface" will asymptotically approach its "Schwarzschild radius", but without never crossing it:

$$r_\sigma=2\mu_\sigma+r_\sigma^\Delta\approx 2\mu_\sigma+e^{\frac{A(2\mu_\sigma)}{2\mu_\sigma}}e^{\frac{-t}{2\mu_\sigma}} \quad \text{(eq. 11)}$$

As the value of $r_\sigma$ is extremely closed to $2\mu_\sigma$, the value of $\mu_\sigma$ may be replaced by $r_\sigma/2$ in the functions which depend on $\mu_\sigma$ :

$$2\mu_\sigma \approx r_\sigma \quad \text{(eq. 12)}$$

$$r_\sigma^\Delta \approx e^{\frac{A(r_\sigma)}{r_\sigma}} e^{\frac{-t}{r_\sigma}} \quad \text{(eq. 13)}$$

$$2\mu_\sigma = r_\sigma - r_\sigma^\Delta \approx r_\sigma - e^{\frac{A(r_\sigma)}{r_\sigma}} e^{\frac{-t}{r_\sigma}} \quad \text{(eq. 14)}$$

As this expression is valid for any "σ-surface", we have finally got an approximate expression for the dependence of $\mu$ on radius and time:

$$\mu_{(r,t)} = \frac{1}{2} \cdot \left[ r - e^{\frac{A(r)}{r}} e^{\frac{-t}{r}} \right] \quad \text{(eq. 15)}$$

If we replace the previous expression into the initial "Schwarzschild-like" metric, we will finally obtain the following metric:

$$ds^2 = -h_{(r,t)} dt^2 + h_{(r,t)}^{-1} dr^2 + r^2 d\Omega^2 \quad \text{(eq. 16a)}$$

$$h_{(r,t)} = h_{(r,t)}^{outside} \cdot H(r - r_{edge}) + h_{(r,t)}^{inside} \cdot H(r_{edge} - r) \quad \text{(eq. 16b)}$$

$$h_{(r,t)}^{outside} = 1 - \frac{2M}{r} \quad \text{(eq. 16c)}$$

$$h_{(r,t)}^{inside} = e^{f(r) - \frac{t}{r}} \quad \text{(eq. 16d)}$$

$$r_{edge} = 2M \cdot \left[ 1 + e^{f(2M) - \frac{t}{2M}} \right] \quad \text{(eq. 16e)},$$

where we have defined $f(x) = \frac{A(x)}{x} - \ln x$ , and where $H(x)$ stands for the Heaviside step function. "M" corresponds to the total energy-mass of the collapsing body.

We have finally got a suitable expression for the gravitational collapse. Contrarily to Schwarzschild's original metric, this one has no singularity (or pseudo-singularity) at a finite radius (it may be divergent at radius = 0, but this point contains a null measure of mass, and its physical relevance is expected not to be significant). Also, in contrast with Schwarzschild's original metric, this metric is not stationary, but it is continuously evolving with time.

In spite of all simplifications, all the essential physics of the problem are contained in equations (at least from the mechanical point of view, *See Appendix B*), and there is no reason to think that qualitatively important differences should be observed in a more complete model of gravitational collapse (*See Appendix C*).

## IV. DISCUSSION

In our toy model, we have obtained a metric in which we can remark the following features:

1- Outside the region of the space where the mass of the collapsing star is located, the metric is strictly identical to Schwarzschild's one (certainly, it is in this way *by construction*)

2- Inside the region of the space where the mass of the collapsing star is located, every element of mass is asymptotically traveling towards the surface where an event horizon would be formed if that amount of mass crossed it. As it never crosses it, no event horizon is ever formed, but the space-time deformation is more and more intense at every point as collapse progresses (*See Fig. 2*).

3- The limit of the region mass, if no mass is added or lost to the total mass of the collapsing star (which we have called M), also asymptotically approaches the surface which would become the event horizon if the whole of mass crossed it.

Therefore, we have got to describe the gravitational collapse in terms of a well-behaved non-singular metric.

In respect to the value of mass M, there is an important fact to take into account: it will be a function of time. Certainly, it is expectable that for long periods its rhythm of variation would be slow enough to consider it a constant at practical effects, but in determinate moments of its natural evolution the collapsing star may receive greats amounts of additional mass, for instance when another star approaches it. Nonetheless, in absence of abrupt supplies of matter, the mass of the collapsing star would be softly but continuously increasing or decreasing in function of the preeminence of two opposed processes: the intake of cosmic background radiation, and the emission of thermal (or non-thermal) radiation.

When the emission of radiation would be the preeminent process, all the mass of the collapsing star would be progressively emitted, at a very slow rate, but in a finite period of time. As a matter of fact, the rhythm of thermal emission of a collapsing star should be reviewed, as it is not a black hole, but at the moment we will assume as a hypothesis a similar thermodynamical behaviour for both. Under this assumption, the power of emission of the body is inversely proportional to the square power of mass.

If the macroscopic thermodynamical behavior may be similar (and we point out again that there are not evident strong objections to *a priori* reject this hypothesis), the microscopic emission process must be radically different in respect the established model. That is an obvious consequence of the fact that there is no event horizon in our new paradigm. Radiation and particle emission should be therefore supposed to be emitted in the same way that they are emitted by any body at a certain temperature. Thus, the conservation of information becomes a trivial subject, and there is no further need to conjecture the propagation of mass-negative particles.

There is another significant consideration to take into account in respect to thermal radiation: in our "Onion-Shell" model it should be expected that not only the whole of the collapsing star emitted radiation towards the environment, but also that the inner shells emitted radiation towards the exterior ones. That would imply that for any given radius $r$ the space-temporal deformation may not get an arbitrarily great value but a limited one (*See Appendix B*). It can be easily seen how the outer shells of the collapsing star would be the first ones to be evaporated.

Nonetheless, as we have mentioned, there exist a process which will behave in an antagonistic way in respect to the emission of thermal radiation by the collapsing star, and that is the intake of cosmic background radiation. If the emission of thermal radiation is greater than the assimilation of cosmic radiation, the collapsing star will vanish. If the assimilation of cosmic radiation is greater than the emission of thermal radiation, the collapsing star will increase its mass, so that the additional mass will be added as outer shells to the previous ones. Thermal radiation depends on the mass of the collapsing star, and the cosmic background radiation depends on the scale factor of the Universe. As the Universe seems to be expanding in an accelerated way, the density of cosmic background radiation is expected to get arbitrarily small values after a long enough period of time. It could be discussed if all the collapsing stars will finally vanish or if some may eventually keep on growing for all the History of Time. In spite of this fact, even when never vanishing, in no moment of the evolution of the star could it be properly stated for it to be a black hole. The term "Black Hole" is no longer a good one for referring the final stage of a massive star, for the star is never completely collapsed. It perhaps could be suitable the denomination of "grey hollow" (not so black as black, and not so a hole as a hole).

Investigation in the field should explore the several aspects which this paper mentions but which are not exhaustively developed, as the corrections which should be performed in the proposed metric in order to rightly describe a slower collapse (not a collapse at the speed of light; *See Appendix C*), to determine realistic expressions for the parameters corresponding to the contour conditions, to check the thermodynamical expressions for entropy and temperature, and to predict the fate of the collapsing stars in function of the temporal dependence of the Universe scale factor.

## V. CONCLUSIONS

Strong principles as the invariance of causality make necessary the review of the established model of Black Hole. Several inconsistent aspects have driven us to substitute it by the new paradigm of Never-Stationary Gravitational Collapse (or collapse in "Onion Shells"), with non singular metrics. In the new paradigm, event horizons are never formed, and the "paradox of information" is clearly solved.

## VI. FIGURES

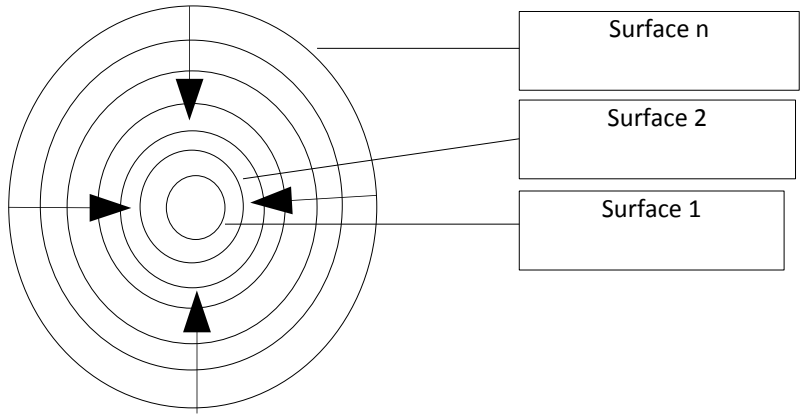

**Fig. 1.** In order to study the evolution of the collapsing body, it has been divided into a series of surfaces such that each one contains a fixed quantity of mass inside it. We assume purely radial movement for every element of mass. Therefore, the radius of every surface changes with time, as those must be co-mobile with matter, but the spherical shape is preserved.

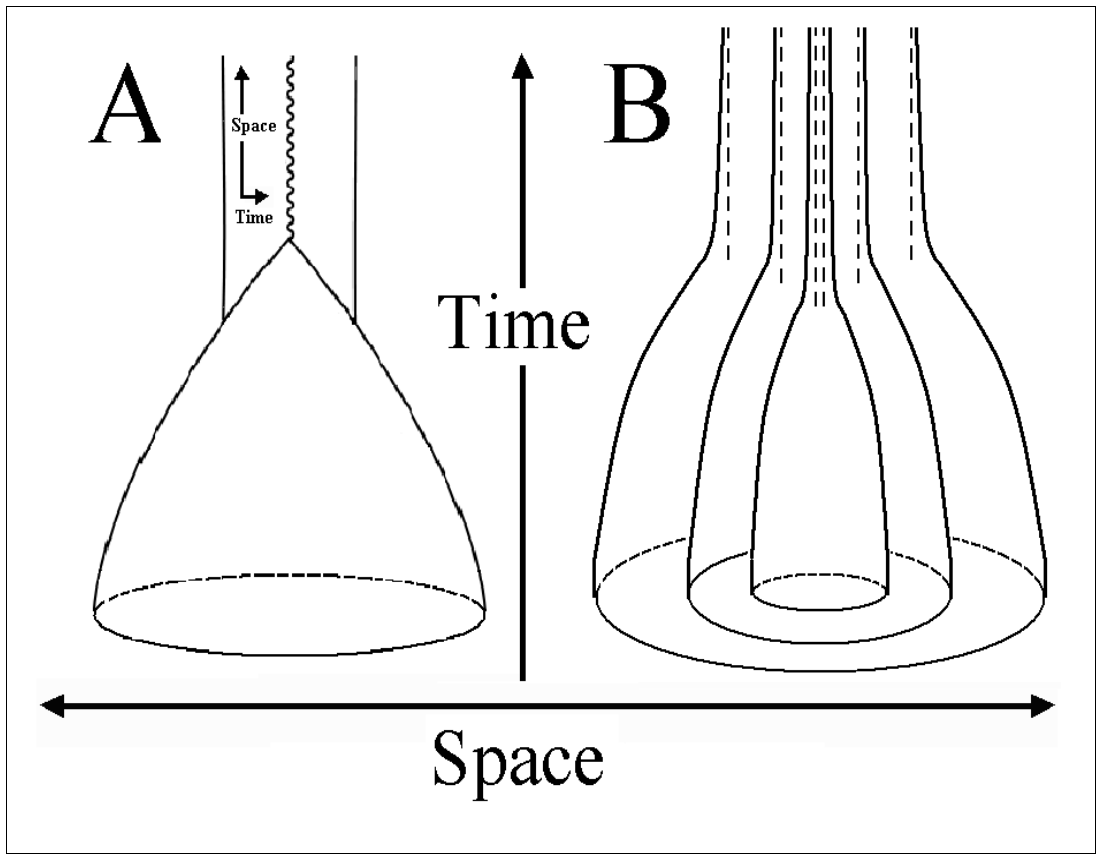

**Fig 2.** Diagram A shows the established model of gravitational collapse, the standard Black Hole, with all the matter getting a central singularity and an event horizon being formed at the Schwarzschild radius; inside the event horizon, time and space coordinates are interchanged. Diagram B shows the Never-Stationary Gravitational Collapse model (or collapse in "Onion Shells"), in which an event horizon tends to be formed at every point of the interior of the collapsing star; every "shell" of mass approaches asymptotically the surface which would become an even horizon if it were actually crossed.

**VI.VII. APPENDICES**

*Appendix A:*

**Demonstration that the perspective of an external observer concerning collapse may not be a simple optical effect**

Let's assume that we are given a stationary metric of the following form:

$$ds^2 = -g_{tt}(x,y,z)dt^2 + g_{ij}(x,y,z)dx^i dx^j \quad \text{(eq. A1)},$$

where $i,j = 1,2,3$ and where $x^1 = x, x^2 = y, x^3 = z$. It is obvious that this metric is completely symmetrical in respect to temporal inversion:

$$t \to -t \Rightarrow ds^2 \to ds'^2 = -g_{tt}(x,y,z)d(-t)^2 + g_{ij}(x,y,z)dx^i dx^j; d(-t)^2 = (-dt)^2 = dt^2; ds'^2 = ds^2$$
(eq. A2)

Therefore, in any situation where a metric of the form given in equation "eq. A1" is given (and certainly Schwarzshild's metric is included in this group), every trajectory in space must be necessarily reversible. Therefore, if an object entered a black hole but the light could not escape from it, it would imply that the object would have traveled at a speed greater than the speed of light.

Furthermore, we have at the same time proven that there exist only two logical possibilities concerning real physical stationary situations:

a) Event horizons do not exist at all in any real physical stationary situation
b) Event horizons may exist in stationary physical situations, but they can never crossed (not in one sense nor in the opposed one)

*Appendix B:*

**Thermal corrections to Never-Stationary Collapse metric**

In order to derive the metric of the gravitational collapse, we have considered mobile surfaces containing a fixed quantity of mass. We have therefore deduced that every surface approached asymptotically a critical radius $r_\mu = 2\mu$.

Now we are going to consider surfaces of fixed radius instead of surfaces of fixed mass. We will define the quantity $\mu_\Sigma^\Delta$ as the amount of mass that should cross the considered $\Sigma$-surface of constant radius $r_\Sigma$ in order that an event horizon were formed, that is, $\mu_\Sigma^\Delta = \frac{r_\Sigma}{2} - \mu_\Sigma$, being $\mu_\Sigma$ the amount of mass actually contained inside the surface. Given the lineal dependence between mass and radius in a collapsing star, it can be easily noticed that in the same way that the variation of $r_\sigma^\Delta$ were exponential when considering $\sigma$-surfaces of fixed mass, the variation of $\mu_\Sigma^\Delta$ is equally exponential when considering $\Sigma$-surfaces of constant radius:

$$\mu_\Sigma^\Delta = \mu_{\Sigma,0}^\Delta \cdot e^{\frac{-t}{r_\Sigma}} \quad \text{(eq. B1)}$$

Differentiating in respect to time, we have:

$$\frac{d\mu_\Sigma^\Delta}{dt} = \frac{-\mu_\Sigma^\Delta}{r_\Sigma} \quad \text{(eq. B2)}$$

To the previous differential expression, obtained according only to "mechanical" considerations, we should add a "thermal" term according to the fact that the mass $\mu_\Sigma$ contained inside the Σ-surface will emit thermal radiation to the exterior of the surface. We will assume a similar thermal behavior to that proposed by Hawking for the radiation of black holes, that is, inversely proportional to the to square of the mass [6, 7]. By adding the thermal term, we obtain the following differential equation:

$$\frac{d\mu_\Sigma^\Delta}{dt} = \frac{-\mu_\Sigma^\Delta}{r_\Sigma} + \frac{\alpha}{\mu_\Sigma^2} \quad \text{(eq. B3)},$$

where $\alpha$ is a constant (and whose absolute value is considerably small).

In the previous equation, $\mu_\Sigma$ may be substituted without problem by $\frac{r_\Sigma}{2}$, as the difference between both values is simply differential. Then, we get:

$$\frac{d\mu_\Sigma^\Delta}{dt} = \frac{-\mu_\Sigma^\Delta}{r_\Sigma} + \frac{4\alpha}{r_\Sigma^2} \quad \text{(eq. B4)}$$

In order to solve the equation B4, we will separate the mass $\mu_\Sigma^\Delta$ into a constant term $\mu_{\Sigma,eq}^\Delta$ and a temporal dependent term $\mu_{\Sigma,t}^\Delta$, that is, $\mu_\Sigma^\Delta = \mu_{\Sigma,eq}^\Delta + \mu_{\Sigma,t}^\Delta$. Therefore, we can separate equation B4 into equations B4a and B4b:

$$0 = \frac{-\mu_{\Sigma,eq}^\Delta}{r_\Sigma} + \frac{4\alpha}{r_\Sigma^2} \quad \text{(eq. B4a)}$$

$$\frac{d\mu_{\Sigma,t}^\Delta}{dt} = \frac{-\mu_{\Sigma,t}^\Delta}{r_\Sigma} \quad \text{(eq. B4b)}$$

By solving both equations, we obtain:

$$\mu_{\Sigma,eq}^\Delta = \frac{4\alpha}{r_\Sigma} \quad \text{(eq. B5)}$$

$$\mu_{\Sigma,t}^\Delta = (\mu_{\Sigma,0}^\Delta - \mu_{\Sigma,eq}^\Delta) \cdot e^{\frac{-t}{r_\Sigma}} \quad \text{(eq. B6)}$$

$$\mu_\Sigma^\Delta = \left(\mu_{\Sigma,0}^\Delta - \frac{4\alpha}{r_\Sigma}\right) \cdot e^{\frac{-t}{r_\Sigma}} + \frac{4\alpha}{r_\Sigma} \quad \text{(eq. B7)}$$

If we have into account that the initial "deficit of mass" $\mu_0^\Delta(r)$ must be a positive well behaved function, for instance $\mu_0^\Delta(r) = \frac{r}{2} e^{f(r)}$, we obtain the following expression for the "deficit of mass" $\mu^\Delta(r,t)$:

$$\mu^\Delta(r,t) = \left(\frac{r}{2} e^{f(r)} - \frac{4\alpha}{r}\right) \cdot e^{\frac{-t}{r}} + \frac{4\alpha}{r} \quad \text{(eq. B8)}$$

This fact implies that the function $h_{(r,t)}^{inside}$ (defined in equations "eq. 16a-e") should be thermally corrected in the following way:

$$h_{(r,t)} = \left(e^{f(r)} - \frac{8\alpha}{r^2}\right) \cdot e^{\frac{-t}{r}} + \frac{8\alpha}{r^2} \quad \text{(eq. B9)}$$

*Appendix C:*

**A more general expression for the metric of the gravitational collapse with spherical symmetry**

In order to simplify notation, in equations "eq. 16a-e" (*See the Toy Model section*) we have defined the funcion $h_{(r,t)}$, which fully characterizes the metric of a spherically symmetric gravitational collapse, and we have divided it in the following way:

$$h_{(r,t)} = h_{(r,t)}^{inside} \cdot H(r_{edge}(t) - r) + h_{(r,t)}^{outside} \cdot H(r - r_{edge}(t)) \quad \text{(eq. C1)}$$

$$h_{(r,t)}^{outside} = h_{(r,t)}^{Schwarzshcild's} = 1 - \frac{2M}{r} \quad \text{(eq. C2),}$$

where $r_{edge}(t)$ corresponds to the radius of the area of space where the mass of the collapsing star is distributed, and whose value must logically correspond to the following one:

$$r_{edge}(t) = 2M \cdot \left[1 + h_{(2M,t)}^{inside}\right] \quad \text{(eq. C3)}$$

Outside $r_{edge}(t)$, the metric of the collapsing star coincides exactly with Schwarzschild's metric, as equation "eq. C2" indicates. According to "eq. 16d", the expression for $h_{(r,t)}^{inside}$ corresponds to the following one:

$$h_{(r,t)}^{inside} = e^{f(r)} \cdot e^{\frac{-t}{r}} \quad \text{(eq. C4),}$$

where $e^{f(r)}$ depends on the contour conditions of the collapse.

If we check out the assumptions which we have performed in order to simplify our calculations, it is straightforward to notice that the expression of $h_{(r,t)}^{inside}$ may be easily extended to less restrictive conditions and to a higher degree of precision in the following way:

$$h_{(r,t)}^{inside} = \left[e^f(r) - \frac{2}{r} \cdot P\left(\frac{r}{2}\right)\right] \cdot e^{\frac{-B^{(r,t)}t}{r}} + \frac{2}{r} \cdot P\left(\frac{r}{2}\right) \quad \text{(eq. C5),}$$

where:

(a) $P(\mu)$ corresponds to the power of thermal radiation of a collapsing body with mass $\mu$ (*See Appendix B*)

(b) $B^{(r,t)}$ is the correction factor which has into account the fact that the collapse is not taking at the speed of light, but to a lesser speed, which we will call $\beta_\sigma(t)$. As a matter of fact, $\beta_\sigma(t)$ is expected to be strictly increasing with time, but its value will be always comprised between 0 and 1. The subindex $\sigma$ indicates that in the general case the speed of collapse may differ from a collapsing surface to another one. The connection between $\beta_\sigma(t)$ and $B^{(r,t)}$ may be easily determined:

$$B_\sigma(t) = \frac{\int_0^t \beta_\sigma(t')\,dt'}{t} \quad \text{(eq. C6)}$$

$$0 < \beta_\sigma(t) < 1 \,\forall\, t \Rightarrow 0 < B_\sigma(t) < 1 \,\forall\, t \quad \text{(eq. C7)}$$

$$B_\sigma(t) = B^{(r_\sigma,t)} \quad \text{(eq. C8)}$$

The equation C4 keeps on being inadequate for the initial moments of collapse, for the we have conserved the approximation of considering $|r_\sigma^\Delta| \ll |\ln r_\sigma^\Delta|$, which is valid for small values of $r_\sigma^\Delta$.

## VIII. Acknowledgements

We want to thank Mr. Jordi Busqué Pérez to aim us to write this article, and Mr. Juan Ortega Pérez to help us with the orthographic revision.